Connecting Three Pivotal Concepts in K-12 Science State Standards and Maps of Conceptual Growth to Research in Physics Education

Chandralekha Singh[1] and Christian D. Schunn[2,3]

Department of Physics and Astronomy[1], Department of Psychology[2] and Learning Research and Development Center[3]

University of Pittsburgh, Pittsburgh, PA 15260

Abstract

This paper describes three conceptual areas in physics that are particularly important targets for educational interventions in K-12 science. These conceptual areas are force and motion, conservation of energy, and geometrical optics, which were prominent in the US national and four US state standards that we examined. The four US state standards that were analyzed to explore the extent to which the K-12 science standards differ in different states were selected to include states in different geographic regions and of different sizes. The three conceptual areas that were common to all the four state standards are conceptual building blocks for other science concepts covered in the K-12 curriculum. Since these three areas have been found to be ripe with deep student misconceptions that are resilient to conventional physics instruction, the nature of difficulties in these areas is described in some depth, along with pointers towards approaches that have met with some success in each conceptual area.

## *Introduction*

Connecting the K-12 science standards and maps of conceptual growth to research on common difficulties and strategies for helping students develop a good grasp of the pivotal concepts is critical for ensuring that our K-12 students master the concepts. This connection between the standards and research on student difficulties in learning the concepts can help all stakeholders including teachers who can incorporate them in instruction, and science faculty members planning professional development activities for K-12 teachers because they may not necessarily know the links between different conceptual areas of science and the standards.

Unfortunately, K-12 science curricula have often been described as being a mile wide and an inch deep (Frelindich, 1998), leaving students with little understanding of or interest in science. The problem is further intensified because many elementary teachers are teaching science with little background in science, and many middle school and high school science teachers are teaching out of field (Ingersoll, 2003; Shugart & Houshell, 1995), or perhaps with out-of-date knowledge (Griffith & Brem, 2004). Thus, it is very difficult to provide good professional development for science teachers on so many different science topics.

One possible solution is to emphasize fewer topics. Indeed, the AAAS Project 2061 Benchmarks for Science focus on a smaller set of coherent themes that are typically covered in many K-12 science courses. There are many benefits of having a smaller set of topics to teach: science education researchers can focus their research efforts to analyze and understand the learning issues on a more focused set of concepts; science curriculum developers can develop curriculum with greater research support and more focused testing; faculty members involved in teacher preparation can focus their in-service and pre-service professional development activities on thoughtfully prepared and tested strategies; teachers can spend time exploring the interplay of science processes and science content with their students rather than racing through a textbook of science facts and stories; and students can come to deeply understand and appreciate science as a way of thinking and interacting with the world around them (Lederman, 1992).

Unfortunately, the majority of the state science standards in the US have much broader content coverage than the AAAS Benchmarks for Science. The current climate for K-12 science education in the US is one of high stakes accountability under the No Child Left Behind legislation. Because performance on state standardized test is a key variable, and because the tests focus solely on broad state-specific standards, the pressure on students, science teachers, school districts, schools of education, and curriculum developers continues to be in the direction of breadth of coverage.

Despite such pressure, there is room in the K-12 science curriculum for higher quality science experiences that can help students develop problem solving and reasoning abilities. There are some foundational science concepts that have more overall influence on student performance than others, and high quality experiences could be created to enable the learning of these concepts. Some research-based materials that provide such experiences have already been created. It is their effective implementation in K-12 education that remains problematic. The focus of the current paper is to explore this conjecture in the context of physics. Specifically we ask whether there are a set of physics concepts that are widely found in state standards, are foundational for later learning of other K-12 science concepts, and are traditionally very difficult to learn.

With such information in hand, faculty members involved in teacher preparation, curriculum developers, and teachers could be better informed about what physics concepts are worthy of extended inquiry which is a key decision when using inquiry-based approaches for improving students' learning. Science teachers who are typically required to update their knowledge with ongoing professional development (Fishman, Marx, Best, & Tal, 2003) will also find this paper useful. This paper tries to capture the core K-12 learning challenges of physics, bridging the often disparate worlds of high stakes accountability, deep science disciplinary perspectives, and learning challenges.

### *Analyzing State Standards with a Focus on Physics*

From our analysis of standards and curricula in the US, physics and chemistry are usually treated together through the elementary years under the label of physical sciences, and typically with considerably less emphasis than the coverage devoted to biology and earth science concepts. In the middle school years, physics and chemistry emerge as separate but related disciplines. In high school, physics and chemistry are treated as entirely disconnected, although to physicists, the same underlying physics concepts can be found in high school chemistry, biology, and earth science courses (e.g., conservation of energy, forces in equilibrium).

In this paper, we present a three-part analysis of the conceptual landscape in K-12 physics. In the first part of the analysis, we examined concept maps—some from the Science Atlas created by Project 2061 and some developed by us when they were not available in the Science Atlas—of different conceptual clusters that plot how physics concepts in the K-12 curriculum are related to one another. We looked for concepts that were pivotal nodes within the maps. In other words, we looked for concepts that were foundational to many other related concepts. Since the structure of physics is very hierarchical, there are deep connections within K-12 physics, with cross-connections between sub-areas of physics (e.g., between forces and motion, conservation of energy, and electricity and magnetism). Similarly there are important connections and bridges to other K-12 sciences. Without engaging in scientific reductionism, one can note that all of the concepts that are shared across the K-12 sciences (except for the process ideas) are essentially physics concepts (e.g., conservation of energy).

In the second part of the analysis, we examined state science standards from four states representing a wide range of state standards. With only 4 states, one cannot be exhaustive, but we tried to cover the following dimensions: very large, very small, and mid-sized states (reflecting differential resources in the construction of standards); and West, Central, and East states (reflecting different values from historical populations and industries). But most importantly, we tried to cover states that had very different styles of standards. The states we selected and their standard style included: California (extremely detailed, very fact oriented, organized by grade level), Colorado (mostly conceptual, organized by discipline and grade groups 4-8-12), Rhode Island (moderately detailed on a more select set of concepts, based on Project 2061, organized by themes and grade groups 2-5-8-12), and Wisconsin (extremely conceptual, organized by discipline, grade 4-8-12, and theme). We looked for concepts that were prominently found (i.e., as full standards on their own, rather than buried as one minor example in another standard) in the science standards for all four states, and at the same approximate level (e.g., at the middle school level).

It should be noted that physics is the oldest and most basic science, and thus one may expect the topics for inclusion into K-12 physics courses to be relatively stable. Indeed, physics K-12 content involves mostly scientific work from over 100 years ago, and not for historical reasons but rather because the core classical physics knowledge has not seen much change. By contrast biology has seen an explosion in the amount of knowledge known in the last 20 years, e.g., knowledge related to the human genome, and these changes are reflected in the curriculum. Interestingly, even in physics, there is only moderate agreement across state standards in content coverage. Some big ideas (e.g., magnetism) are found in elementary standards in one state and in high school standards in another state. Some big ideas are completely absent in some state standards. For example, electricity concepts are not universally found in state standards.

In the third part of the analysis, we examined the research literature on difficulties in learning physics to determine why pivotal physics concepts in the state standards are challenging for students to learn and research-based strategies that have been found successful.

The physics at the high school level demands a certain level of mathematical sophistication and quantitative expertise in at least algebra and trigonometry to avoid cognitive overload (Larkin, McDermott, Simon & Simon, 1980; Singh, 2002; Singh, 2008b). The mathematics in physics often represents a serious challenge for many students (Reif, 1981; Larkin & Reif, 1979; Singh, 2004). However, the third part of our analysis focused on conceptual difficulties in learning physics. Regardless of how proficient students are in quantitative analysis, conceptual understanding is necessary to be able to perform quantitative analysis beyond guessing or "plug and chug" (Mazur,

1997; Kim & Pak, 2002; McDermott, 2001; Singh, 2008a, 2008c). Research shows that even honors students have conceptual difficulties in learning physics (e.g., difficulty in distinguishing between displacement, velocity and acceleration) similar to the general student population (Peters, 1982).

Finally, we sought those physics concepts that were salient in all three steps: conceptually pivotal, found in all four state standards, and particularly difficult to learn. Three concepts emerged: Newton's laws (qualitatively only at the middle school level or qualitative and quantitative at the high school level), conservation of energy (at the high school level) and geometrical optics (at middle and high school levels). No other concepts came close to meeting all three criteria.

The remainder of this paper presents the case for each of these three concepts. Each section begins a discussion of the role of the identified concept in the broader conceptual landscape. Second, there is a brief discussion of how state standards talk about the concept and at what level (high school or middle school) the concept can be commonly found. Third, there is an in-depth discussion of what makes that particular concept difficult to learn, as a resource for teachers, those involved in professional development, and curriculum developers. Finally, there is brief mention of approaches that have seen some success in teaching the particular concept.

### *Newton's Laws*

Force and motion are fundamental concepts in all sciences and are related to diverse physical phenomena in everyday experience. These concepts provide the backbone on which many other science concepts are developed. According to the Atlas of Science Literacy Project 2061 Motion maps (see Appendix A), children in grades K-2 should be given an opportunity to learn about various types of motion e.g., straight, zigzag, round and round, back and forth, fast and slow and how giving something a push or a pull can change the motion. The map shows a gradual transition to helping students develop more sophisticated ways of thinking about forces and motion in later grades. For example, children in grades 3-5 should be taught how forces cause changes in the speed or direction of motion of an object and a greater force will lead to a larger change in these quantities. Children in 6-8 grades should learn Newton's laws, relative velocity concepts, and their implication for motion with a central force (e.g., planetary motion) mostly qualitatively while those in grades 9-12 should learn these concepts more elaborately and quantitatively.

In the map in Appendix A, the concepts that are a component of Newton's laws are indicated in italics. In the middle grades, there is a recommended emphasis on a qualitative understanding of Newton's laws, followed by a quantitative understanding in high school. It is important to note that the qualitative understanding of Newton's laws, and to some extent the quantitative understanding of Newton's laws is the foundation of many other related concepts.

Turning to the state science standards, one finds that only Newton's second law ($F=ma$), of all force and motion concepts, is found consistently in the standards. Table A1 presents the relevant state science standards. At the middle school level, the required understanding is very qualitative, and thus the language does not directly refer to the law itself. It is interesting to note that in the Colorado and Wisconsin standards, the language in the standards is so general for the relevant middle school standards that a variety of force and motion concepts at the qualitative level are invoked, and only a person very knowledgeable in physics is likely to realize that Newton's second law is highly relevant here.

At the high school level, the relevant science standards are much more quantitative and specific to Newton's second law, although only the California standards have the actual equation and name the law specifically. Rhode Island standards describe the key quantitative relationship in the law in words rather than in an equation. Colorado and Wisconsin standards again use very abstract terms such that only a person very knowledgeable in physics would realize that Newton's laws were being invoked.

The standards particularly emphasize Newton's second law. However, since all the three laws of motion are intertwined, an understanding of all the three laws of motion is necessary for a good understanding of force and motion. Therefore, we will discuss all the three laws of motion in some detail.

Unfortunately, the teaching of force and motion concepts is quite challenging (Camp & Clement, 1994; Champagne, Klopfer & Anderson, 1980; Clement, 1983; Halloun & Hestenes, 1985a, Halloun & Hestenes, 1985b; McDermott, 1984; McDermott, 2001; Singh, 2007). Students are not blank slates. They constantly try to make sense of the world around them. Since force and motion concepts are encountered frequently in everyday experiences, people try to rationalize their experiences based upon their prior knowledge, even without formal instruction. According to Simon's theory of bounded rationality (Simon, 1983; Simon & Kaplan, 1989), when rationalizing the cause for a phenomenon, people only contemplate a few possibilities that do not cause a cognitive overload and appear consistent with their experience. Accordingly, students build "micro" knowledge structures about force and motion that appears locally consistent to them but are not globally consistent. These locally consistent naive theories due to mis-encoding and inappropriate transfer of observation are termed "facets" by Minstrell (1992) and "phenomenological primitives" by diSessa (Smith, diSessa & Roschelle, 1993).

Cognitive theory suggests that preconceptions and difficulties about a certain concept are not as varied as one may imagine because most people's everyday experiences and sense-making is very similar (Reason, 1990; Tversky & Kahneman, 1974). Therefore, regardless of the grade-level in which force and motion concepts are taught, most students have similar preconceptions about motion and forces (Camp & Clement, 1994; Champagne, Klopfer & Anderson, 1980; Clement, 1983; Halloun & Hestenes, 1985a, Halloun & Hestenes, 1985b; McDermott, 1984; McDermott, 2001; Singh, 2007). For example, contrary to the Newtonian view, a majority of students believe that motion implies force and an object moving at a constant velocity must have a net force acting on it. This is an over-generalization of the everyday observation that if an object is at rest, a force is required to set it in motion. Due to the presence of frictional forces in everyday life, such preconceptions are reinforced further, e.g., in order to make a car or a box move at a constant velocity on a horizontal surface one needs to apply a force to counteract the frictional forces. These observations are often interpreted to mean that there is a net force required to keep an object in motion. Research has shown that these preconceptions are very robust, interfere with learning, and are extremely difficult to change without proper intervention (Arons, 1990; Camp & Clement, 1994; Champagne, Klopfer, & Anderson, 1980; McDermott, 1991; McDermott, 1993). They make the learning of the Newtonian view of force and motion very challenging, and old conceptions often reappear after a short time.

In fact, the concepts of force and motion proved very challenging to early scientists prior to Newton and Galileo. Halloun and Hestenes (1985a) discuss how the great intellectual struggles of the past provide valuable insight into the conceptual difficulties of students learning these concepts. The common sense notion of many beginning students conforms more with the medieval Impetus theory of force and motion, than with the Aristotelian view (Halloun & Hestenes, 1985a, 1985b). Students who hold the impetus view tend to believe that if a baseball is hit by a bat, the force of the hit is still acting on the ball long after the ball has left contact with the bat and is in the air.

Research has shown that even after instruction, students' views about force and motion is context dependent and many students solve problems using the correct Newtonian principles under certain contexts while choosing non-Newtonian choices under other contexts (Camp & Clement, 1994; Champagne, Klopfer & Anderson, 1980; Clement, 1983; Halloun & Hestenes, 1985a, 1985b; McDermott, 1984; McDermott, 2001; Singh, 2007). For example, students may cite Newton's first law to claim that an object moving at a constant velocity in outer space (where there is nothing but vacuum) has no net force acting on it but claim that there must be a net force on an object moving at a constant velocity on earth. Many students incorrectly believe that Newton's first law cannot hold on earth due to the presence of friction and air-resistance. Similarly, even after instruction in Newton's third law, students have great difficulty recognizing its significance in concrete situations (Halloun & Hestenes, 1985a, Hammer & Elby, 2003, Mazur, 1997). For example, if students are asked a question involving collision of a small car and a big truck after instruction in Newton's third law, a majority believe that the big truck will exert a larger force on the small car. This conception is due to the confusion between force and acceleration (Halloun & Hestenes, 1985a, Hammer & Elby, 2003, Mazur, 1997). Although the magnitude of the force exerted by the big truck on the small car is equal to the magnitude of the force exerted by the small car on big truck, according to Newton's second law, the acceleration of the small car will be more. Therefore, the small car will get damaged more in the collision despite the fact that forces are equal on both car and truck.

Newton's laws are very difficult to teach because there are in fact several distinct preconceptions at play, each of which manifests themselves in many different ways (Halloun & Hestenes 1985a,1985b; Hammer & Elby, 2003, McDermott, 2001). In the sections that follow, we describe some of these difficulties and illustrate the diverse ways in which they manifest themselves.

### *Incorrect Linkage of Force and Velocity Concepts*

Students often confuse velocity and acceleration and believe that the net force on an object is proportional to its velocity. Directly tied to this confusion, students also believe that there must be a force in the direction of motion. Clement (1983) performed a study in which he asked first year college students enrolled in a pre-engineering course to draw a force diagram of a coin just after it has been tossed in the air. A large group of students was asked to draw the diagram on paper before and after instruction while a smaller group was presented with the same task during an interview situation. A common incorrect response was that while the coin is on the way up, the force of the hand must be greater than the gravitational force because the students believed that there must be a force in the direction of motion. Students also claimed that the force of the hand on the coin gradually dies away after the coin is launched, consistent with the "impetus" view. The confusion between velocity and net force also caused students to incorrectly claim that the net force on the coin was zero when the coin was at the highest point and on its way down the gravitational force on it is greater than the force of the hand. This confusion between velocity and net force was pervasive even after instruction. Clement extracted a number of characteristics from student responses and labeled them the "motion implies force" preconception. He noted that students who hold these views believe that a force that "dies out" or "builds up" accounts for changes in an object's speed.

Viennot (1979) used written questions to investigate high school and introductory college students' understanding of force and motion. She posed a problem in which a juggler is playing with six identical balls. At a particular instant the balls were all at the same height from the ground but had velocity vectors pointing in different directions. Students were asked if the forces on all the balls were the same or not. Approximately half of the students noted that the forces are different because of the confusion between velocity and force.

A conceptual standardized multiple-choice assessment instrument that has been used extensively in high schools and introductory college physics courses to evaluate student understanding of and misconceptions related to force and motion concepts is the Force Concept Inventory (FCI) developed by Hestenes et al. (Hestenes, 1995; Hestenes, Wells, & Swackhamer, 1992). A variety of studies with this instrument have shown that a majority of students do not develop a Newtonian view of force and motion after traditional instruction (Hake, 1998).

In one question, an elevator is being lifted up an elevator shaft at a constant speed by a steel cable. In the absence of frictional forces, students are asked to compare the upward force of the cable with the downward force of gravity. According to Newton's first law, both forces should be equal in magnitude since the elevator is moving at a constant velocity. A large number of students believe that the upward force of the cable must have a greater magnitude than the downward force of gravity because the elevator has an upward velocity.

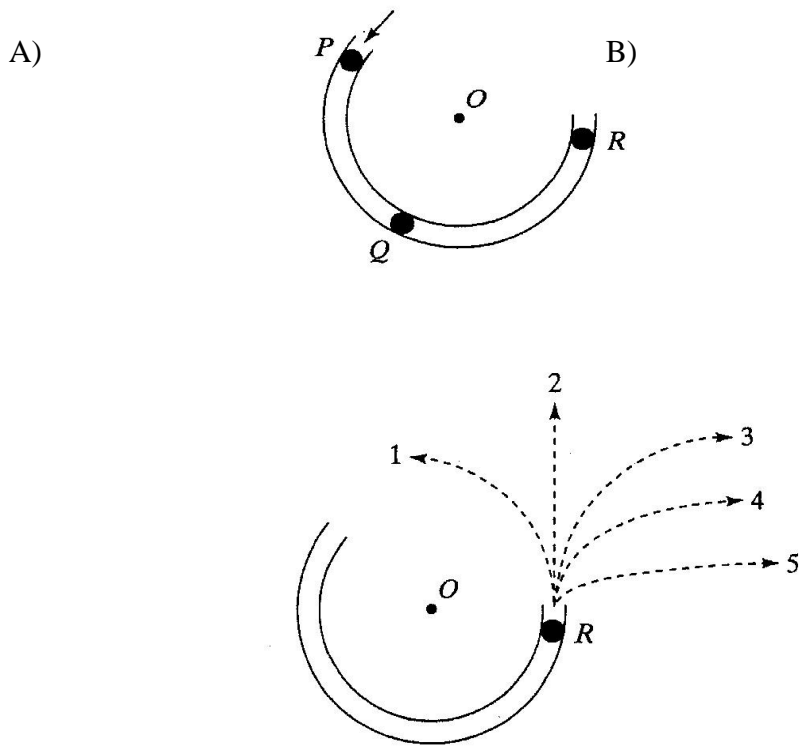

Figure 1. Illustrations of A) a ball at several points within a tube lying on a table, and B) the paths the ball could take on exiting the tube.

Related to this issue, several items on the test probe the misconception that there must be a force in the direction of motion and the forces 'die out" over time. One question on the test has a frictionless channel in the shape of a segment of a circle as shown in Figure 1A. The question notes that the channel has been anchored to a frictionless horizontal tabletop and you are looking down at the table. A ball is shot at high speed into the channel at P and exits at R. Ignoring the forces exerted by air, students are asked to determine which forces are acting on the ball when it is at point Q within the frictionless channel. A very strong distracter is "a force in the direction of motion" which is selected frequently by students. The second part of the question asks for the path of the ball after it exits the channel at R and moves across the frictionless tabletop. According to Newton's first law, the correct response is path (2) as shown in Figure 1B because the net horizontal force on the ball is zero after it exits the channel. The most common distracter consistent with the response to the previous question is path (1) because students believe that there is a force on the ball in the direction of motion that should continue to keep it along the circular path even after it exits the channel.

This bizarre circular conception of impetus is not specific to circular motion in a tube. In another question, students have to predict the path of a steel ball attached to a string that is swung in a circular path in the horizontal plane and then the string suddenly breaks near the ball (Figure 2). A large number of students choose distracter (1) instead of the correct response (2) even after instruction in Newtonian physics.

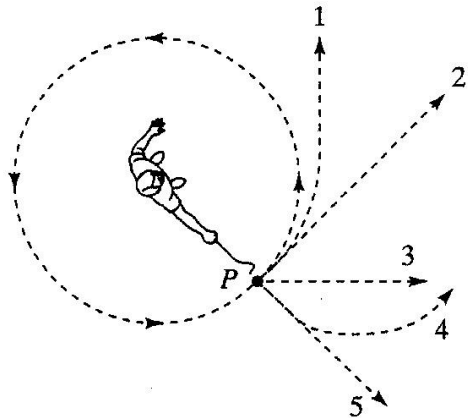

Figure 2. Possible paths taken by a ball swung around on a string and launched at point P.

The problem of impetus conceptions can also be thought of as a confusion between velocity and acceleration. This confusion between velocity and acceleration is illustrated by a question in the FCI in which students are given the position of two blocks at 0.2-second time interval and are asked to compare their accelerations (Figure 3). Neither block has any acceleration because their displacements are equal in equal times. However, a large number of students believe that the block with a larger speed must have a larger acceleration.

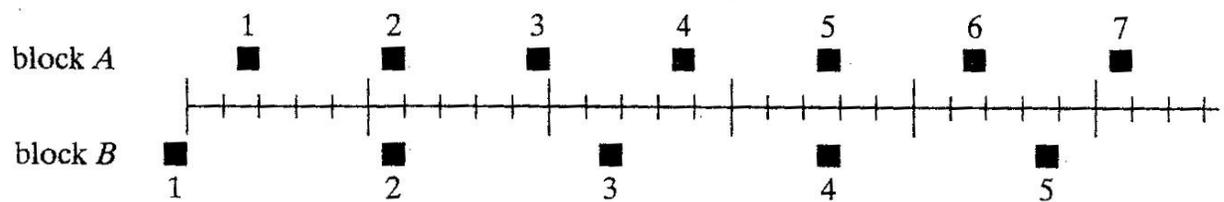

Figure 3. The positions of two blocks taken at 0.2-second intervals.

McDermott et al. (McDermott, 1984; Trowbridge & McDermott, 1981) have shown that in fact the confusion is between displacement, instantaneous velocity (or simply the velocity), and instantaneous acceleration (or simply the acceleration) of an object. One investigation involved asking students to compare the acceleration of two balls sliding down two tracks and whether the accelerations were ever equal for the two balls. About half the students incorrectly claimed that the acceleration would be the same for the two balls at the same point where the velocities of the balls were equal. When the interviewer asked for reasoning, a typical response was that since the acceleration is the change in velocity over time, at the point where the velocity were the same, the rate of change of velocity will be the same as well. Students claimed that since the change in time is the same in both cases, the acceleration at that instant should be the same. This is obviously not correct, because while one can talk about velocity at one position, acceleration is determined by looking at velocity at two different locations (which can be infinitesimally close). In fact, if an object with a zero velocity could not have a non-zero acceleration, the object would never start moving from rest. Part of the difficulty could be due to the confusion between the instantaneous values of velocity and acceleration and their average values for some elapsed time especially for cases where the objects start from rest and are moving in one dimension.

      The impetus misconception also relates to weight/mass confusion. For example, one pervasive naive belief is that the rate at which things fall under the gravitational force is dependent on their weight. There are several items on the FCI that probe this misconception (Hestenes, 1995; Hestenes, et al., 1992). For example, when two balls with different mass are dropped from the same height both balls should take the same time because they both fall under the same gravitational acceleration. A common misconception is that the heavier ball will reach the bottom faster. Another question on the FCI test asks students to compare the horizontal distance from the base of the table covered by metal balls with different masses when they are rolled off a horizontal table

with the same speed. The correct response is that both balls should hit the ground at the same horizontal distance from the table but many students incorrectly believe that the heavier ball will fall horizontally farther due to its greater weight.

Singh has developed several explorations (Singh, 2000; 2002) that greatly improve student understanding of concepts related to force and motion. All the explorations begin by asking students to predict what should happen in a particular situation in which misconceptions are prevalent. For example, the exploration that challenges students' belief that the force of hand still acts on an object after the object is no longer in touch with the hand begins with the following question: "When a baseball soars in the sky after being hit by a bat, the force of the hit still acts on the ball after it has left contact with the bat". Do you agree or disagree with this statement? Explain." After answering this warm-up question, students perform an exploration on a frictionless horizontal air-track. They are asked to push a slider on the track with different initial velocity and then record the velocity and acceleration using a motion sensor and computer. Students are asked to interpret their graph that shows that the velocity is more in the case in which greater initial force was applied to the slider but the acceleration of the slider on the horizontal air-track remains zero for all these cases. They are then asked to interpret what zero acceleration implies about the net force on the object according to Newton's second law. A majority of the students are able to rationalize that the net force on the slider must be zero if the acceleration is zero. They are then asked how this is possible and what it means about the initial force of the hand they applied to make the slider move. Most students are able to interpret that the force of the hand does not act on the slider once it has been let go. Then students are asked to re-evaluate their initial response to the baseball question and whether the force of the hand still acts on it after once it has been let go.

Another exploration helps students understand that the net force is NOT proportional to velocity and helps them distinguish between acceleration, velocity and displacement. Students are given a situation in which two friends are driving in parallel lanes. One person is going at a constant velocity of 30 m/s while another person starts from rest and is accelerating at $1 m/s^2$. They cross at some point. Students are asked about which variables (acceleration, velocity or displacement) are the same for both friends when they cross. Students perform this exploration with sliders on parallel air tracks in which they observe using motion sensors and computer graphs that while displacements are the same, the velocity and acceleration are not the same when the two sliders cross. Students rationalize these observations and learn to make distinction between different variables related to motion. This type of exploration can also be helpful in teaching students about the difference between the instantaneous and average velocities.

Another exploration helps students understand that an object dropped from a moving car or airplane has the same horizontal velocity as the car or airplane. Students start the exploration by answering the following question: Predict whether a ball dropped from your hands while you are standing on a moving walkway at the airport will fall behind you, in front of you, or next to you ignoring air-resistance. Then students perform an exploration with a ball launcher moving at a constant velocity on a horizontal air-track. They find that the ball launched vertically from the launcher follows a parabolic path and falls back in the launcher. They have to interpret what it means about the horizontal velocity of the ball after it is launched and the forces acting on it. After the exploration, a majority of the students are able to explain that the ball dropped from a moving walkway will fall next to the person because the ball has the same horizontal velocity as the person.

### *Difficulty in Understanding the Components of the Net Force*

Students often have difficulty figuring out the individual forces acting on an object. This skill is vital for applying Newton's laws and for appropriately determining the net force in various situations. Minstrell (1982) performed a study investigating high school students' preconceptions about what was keeping a book at rest. Many students drew and labeled diagrams that depicted air pressing in from all sides while others noted that air was mainly pressing down on the book. Some students noted that air pressure was helping gravity hold the book down and some explicitly noted that if air was taken away, the book might drift off. Nearly all students invoked gravity but some students thought of gravitational force as a tendency of an object to go down as opposed to the pull of the earth. Only half the students noted that the table exerts an upward force on the book. For the others, the table was incapable of exerting a force; it was simply in the way. Minstrell's modified instruction, which was reasonably successful, included discussions of an object placed on a helical spring, why the spring compresses and its implications for an upward force on the object by the spring.

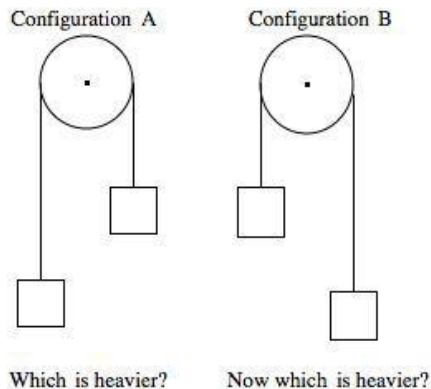

Figure 4. Example Atwood machine configurations with blocks at rest, but with blocks in different locations. Reprinted with permission from Mestre, J. & Tougher, J. Cognitive research--What's in it for physics teachers? The Physics Teacher. 1989, American Association of Physics Teachers.

One common factor involved with difficulties in analyzing the components of a net force is the tendency of beginning students to focus on the surface features of the problem to draw inferences (Mestre & Tougher, 1989, Singh, 2007). The lack of focus on deep features is well illustrated in research involving the Atwood machine, which has two masses connected to each other via a weightless rope as shown in the Figure 4. Research has found that if the rope is lower on one side of the Atwood machine than the other and the whole system is at rest, students predict that the mass on the lower side must be larger (Mestre & Touger, 1989).

In a slightly different version of the set up, researchers clamped the masses at the ends of the rope in the Atwood machine set up, drew students' attention to the fact that the masses on the two sides of the rope were the same (even though they had different sizes) and then asked them to predict what would happen to the masses after un-clamping them. A majority of students predicted that the smaller mass will accelerate downward and the larger mass will accelerate upward. This prediction is in contradiction with the Newtonian analysis in which the net force on each identical mass is zero so there is no acceleration. Therefore, the masses should remain at rest even after the clamp is removed.

There are a variety of techniques that help students correctly analyze the individual components of a net force. Mestre et al. (1989) argue that these kinds of demonstrations, if preceded by the prediction phase, can be powerful tools for creating a state of disequilibrium in students' minds (Ginsberg & Opper, 1969; Gorman, 1972). Following this view, if students are given appropriate guidance and support to assimilate and accommodate Newtonian views about force and motion using free body diagrams, they are likely to be successful (Posner, Strike, Hewson & Gertzog, 1982).

Sokoloff and Thornton (1997) also argue that students are better able to develop Newtonian views of force and motion by preceding lecture demonstrations with a prediction phase. They developed a large number of interactive lecture demonstrations that give students an opportunity to predict the outcome of experiments. The outcomes of these demonstrations often contradict common sense notions and challenge students to resolve the inconsistencies in their prior knowledge and what they observed. Students are then guided through a set of exercises that help them resolve the inconsistencies and build robust knowledge structure. Thornton and Sokoloff have also designed a standardized assessment tool called Force and Motion Conceptual Survey that can be given as a pre- and post-test to assess the extent to which students have developed Newtonian views of force and motion (Thornton & Sokoloff, 1998).

## *Difficulty with the Vector Nature of Variables*

Student difficulty with force and motion is also due to the difficulty with the vector nature of some kinematics and dynamics variables (Aguirre, 1988; Aguirre & Erickson, 1984; Helm & Novak, 1983; Saltiel & Maigrange, 1980). Force, acceleration, velocity and displacement are all vector quantities. Addition of these variables involves knowledge of vector addition and notion of reference frames. In the FCI test, one question asks students about the path of a hockey puck moving horizontally after a force perpendicular to the direction of its velocity is applied to it at an instant. Rather than vectorially accounting for the original velocity, many students believe that the puck will immediately start moving in the direction of the applied force.

McClosley, Caramazza and Green (1980) performed a study in which they asked students who were enrolled in the introductory college physics courses about the path (trajectory) of a pendulum bob after the string was cut when the pendulum was at four different points during its oscillatory motion. Only one fourth of the students provided the correct response for all the four points of the bob. A majority of students ignored the velocity of the bob at the instant the string was cut and 65% noted that the bob would fall straight down (as though it was at rest) when the string was cut at the instant it was passing through its equilibrium position during the oscillatory motion.

A similar misconception is manifested in the FCI test (Hestenes, 1995; Hestenes, et al., 1992). One question in the FCI asks students about the trajectory of a ball dropped from an airplane that is moving horizontally at a constant velocity ignoring air-resistance. Many students do not realize that since the ball is in the airplane when it is dropped, it has the same horizontal velocity as the plane. It should therefore fall along a parabolic path and in the absence of air resistance it should hit the ground right underneath the airplane. Many students believe that the ball would fall behind the airplane because they do not consider its horizontal velocity.

As noted earlier, Singh (2000, 2002) has found that an exploration by students illustrating that a ball launched from a launcher moving on a horizontal track follows a parabolic path and lands back into the launcher can be an effective instructional tool if it is preceded by asking students to make a prediction about the outcome.

In sum, Newton's laws are prominent in US State science standards, are foundational concepts, and are quite difficult for students to learn, for a number of different reasons. We have further identified the kinds of experiences that have been found to help students improve their learning of these concepts.

### *Conservation of Energy*

Similar to the concepts of force and motion, energy is a fundamental concept that is useful in all sciences. The Atlas of Science Literacy Project 2061 does not have a map specifically of energy concepts. Therefore, we created our own map organizing energy concepts found in the National and State Science Standards (see Appendix B). For example, children in grades K-2 should learn about the different forms of energy e.g., sound, light, heat, nuclear, energy of motion at a level consistent with their cognitive development. Children in grades 3-5 should learn at a qualitative level consistent with their expertise that energy cannot be created or destroyed but be converted from one form to another by doing work. Children in grades 6-8 can build on the previous concepts by learning more elaborately about kinetic and potential energies, heat energy and the scientific meaning of "work" in terms of force and distance (for the case where the force and the corresponding motion are in the same direction). High school students can learn these things in greater depth and more quantitatively. For example, in addition to a more in-depth analysis of the previously learned concepts, they can learn about the differences between conservative and non-conservative forces based upon whether the work done by the force depends upon the path, difference between heat and internal energy and how the nuclear energy is harnessed by converting mass into energy using the Einstein's theory of relativity.

      Within this conceptual map, it becomes apparent that the core conservation of energy concepts (indicated in italics) are pivotal in the energy conceptual map in that they support many other energy concepts. Interestingly, the map places some of these notions of conservation of energy as being most appropriate for late elementary and middle school children. However, these concepts are still quite difficult for students in college physics courses, and thus the most important point is that conservation of energy ideas are the foundation of many other energy concepts rather than recommending a particular point at which energy concepts should be taught.

      Of all the energy-related concepts, conservation of energy is the concept found most consistently within the state standards at a given level. All four state standards examined explicitly named conservation of energy at the high school level, while Rhode Island and Wisconsin standards also made some mention of conservation of energy at the middle school level. Also interestingly, neither Rhode Island nor Wisconsin standards refer to a quantitative formulation of conservation of energy ideas, whereas California and Colorado very specifically make reference to quantitative forms of energy conservation and the ability to calculate energy in various forms.

Teaching energy concepts is quite challenging at all levels of instruction (Lawson & McDermott, 1987, Van Heuvelen & Zou, 2001, Singh, 2003). Unlike the concept of force (pull or push), energy concepts are rather abstract and not very intuitive. Due to their abstractness, transfer of learning from one context to another is extremely difficult (Van Heuvelen & Zou, 2001, Singh, 2003). Beginning students often inappropriately categorize problems that can be solved easily using energy concepts because the deep feature of the problem is not discerned. One prevalent hurdle is that the surface features of the target (to which knowledge is to be transferred) do not trigger a recall from memory of the relevant knowledge of energy concepts acquired in a slightly different context. However, research shows that the ability to recognize features based upon deep physical laws improves with expertise. Chi et al. (Chi, Feltovich, & Glaser, 1981; Chi, Glaser, & Rees, 1982) performed a study in which they asked physics experts and introductory physics students to categorize a large number of mechanics problems. While experts characterized them based upon fundamental principles (e.g., Newton's second law, conservation of energy problem etc.), the classification by students was often based upon superficial features (e.g., pulley and inclined plane problem etc.).

Student difficulties with energy concepts have not been investigated as thoroughly as concepts related to force and motion. However, there are investigations that show that effective instruction in energy concepts is quite difficult (Lawson & McDermott, 1987; Singh, 2003; Van Heuvelen & Zou, 2001). Our investigation shows that introductory physics students can get easily distracted by the surface features of the problem and are often unable to employ energy concepts appropriately (Singh, 2003).

We illustrate the factors making learning of and using conservation of energy difficult by drawing heavily on results from a detailed study conducted by Singh (2003). This study designed and administered a research-based 25 item conceptual multiple-choice test about energy and momentum concepts to over a thousand students in several introductory physics courses and conducted individual in-depth interviews with several dozen students using a think-aloud protocol (Chi, 1994, 1997).

### *Difficulty Recognizing a Problem as a Conservation of Energy Problem*

Conservation of energy is very useful for making complex physics problems simple because it allows one to ignore variables whose effects are quite complex and difficult to calculate and combine. For example, many conservation of energy problems allow path traveled to be ignored. But path is very salient to students and is connected to force concepts that have been a constant focus of attention in the classroom. Thus, students focus on calculating forces along a path and fail to recognize that a simpler conservation of energy solution is possible.

Consider the following question from our study related to conservation of energy (Lawson & McDermott, 1987; Singh, 2003; Van Heuvelen & Zou, 2001):

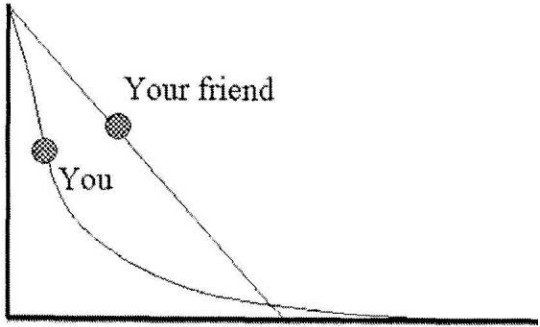

Figure 5. Frictionless slides with the same start height but different shapes. Reprinted with permission from Singh 2003.

1. Two frictionless slides are shaped differently but start at the same height $h$ and end at the same level as shown in Figure 5. You and your friend, who has the same weight as you, slide down from the top on different slides starting from rest. Which one of the following statements best describes who has a larger speed at the bottom of the slide?
    (a) You, because you initially encounter a steeper slope so that there is more opportunity for accelerating.
    (b) You, because you travel a longer distance so that there is more opportunity for accelerating.
    (c) Your friend, because her slide has a constant slope so that she has more opportunity for accelerating.
    (d) Your friend, because she travels a shorter distance so that she can conserve her kinetic energy better.
    (e) Both of you have the same speed.

According to the principle of conservation of mechanical energy, the final speed for both people should be the same. Choices (a) and (c) were the most common distracters. It was clear that many students focused on the surface features of the problem, in particular, the shape of the slides, and did not invoke the principle of conservation of energy.

The exact same kind of results can be found when the objects are in freefall, rather than following the path of a slide. Consider the following pair of problems, illustrating that sometimes students can use conservation of energy in which the change in height is salient, but do not with a nearly identical problem in which the change in height is not salient (Note that students were asked to ignore the retarding effects of friction and air resistance).

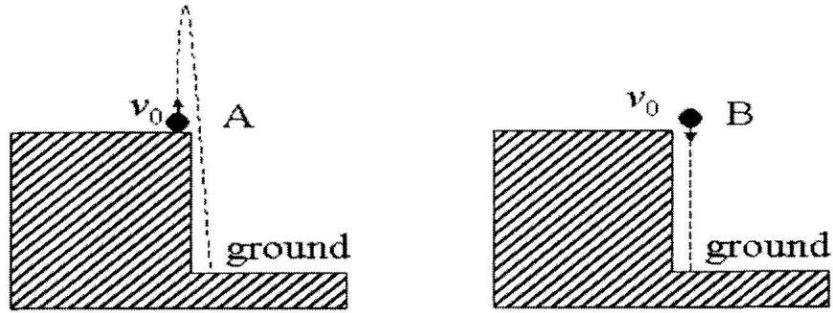

Figure 6. Paths of two identical stones shot with equal initial velocities. Reprinted with permission from Singh (2003).

2. Two identical stones, A and B, are shot from a cliff from the same height and with identical initial speeds $v_0$. Stone A is shot vertically up, and stone B is shot vertically down (see Figure 6). Which one of the following statements best describes which stone has a larger speed right before it hits the ground?
    (a) Both stones have the same speed.
    (b) A, because it travels a longer path
    (c) A, because it takes a longer time
    (d) A, because it travels a longer path and takes a longer time
    (e) B, because no work is done against gravity

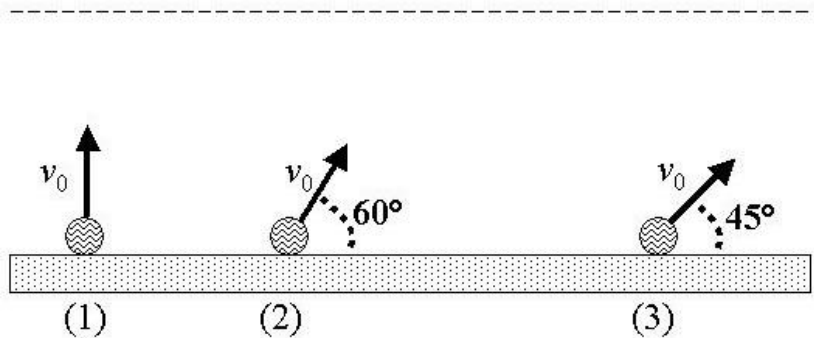

Figure 7. Three balls launched at different angles but with the same initial velocity. Reprinted with permission from Singh (2003).

3. Three balls are launched from the same horizontal level with identical speeds $v_0$ as shown in Figure 7. Ball (1) is launched vertically upward, ball (2) at an angle of $60°$ and ball (3) at an angle of $45°$. In order of decreasing speed (fastest first), rank the speed each one attains when it reaches the level of the dashed horizontal line. All three balls have sufficient speed to reach the dashed line.

(a) (1), (2), (3)
(b) (1), (3), (2)
(c) (3), (2), (1)
(d) They all have the same speed.
(e) Not enough information, their speeds will depend on their masses.

Using the conservation of energy, both stones in problem 2 should have the same speed and all the three balls in problem 3 should have the same speed. Students performed significantly better on problem 2 than problem 3. Problem 2 is similar to the example often presented in the textbooks. In fact, the learning gains between the pre- and post-testing (before and after instruction) was approximately three times larger for problem 2 than problem 3. A majority of students got distracted by the angles that were provided in problem 3 and did not think of using conservation of energy for that problem. Even if relevant knowledge resource about conservation of energy was present in their memory, the superfluous information about angles blocked appropriate association of this problem as a conservation of energy problem. Many started analyzing the problem vectorially, could not go too far, and came to incorrect conclusions.

Conservation of energy problems such as the following that require the student to ignore weight can also be difficult, because many students believe that weight must play a role:

4. While in a playground, you and your niece take turns sliding down a frictionless slide. Your mass is 75 kg while your little niece's mass is only 25 kg. Assume that both of you begin sliding from rest from the same height. Which one of the following statements best describes who has a larger speed at the bottom of the slide?
    (a) Both of you have the same speed at the bottom.
    (b) Your niece, because she is not pressing down against the slide as strongly so her motion is closer to freefall than yours.
    (c) You, because your greater weight causes a greater downward acceleration.
    (d) Your niece, because lighter objects are easier to accelerate.
    (e) You, because you take less time to slide down.

According to the principle of conservation of mechanical energy, the final speed for both people in problem 4 should be the same. Choice (c) was the most common distracter. Here students focused on the weight of the people sliding and did not invoke the principle of conservation of energy.

Problems involving solving for work done that involve conservation of energy also can cause problems. Consider the work done on the blocks in problem 5.

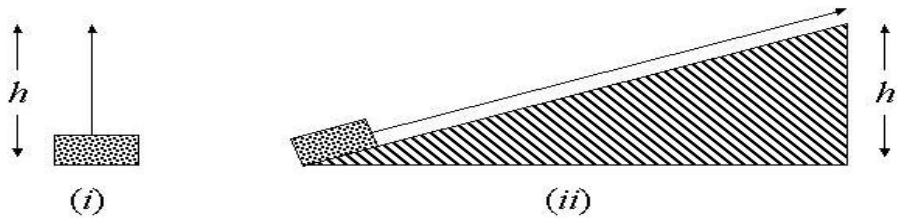

Figure 8. Blocks moved a height h at constant velocity. Reprinted with permission from Singh (2003).

5. You want to lift a heavy block through a height h by attaching a string of negligible mass to it and pulling so that it moves at a constant velocity. You have the choice of lifting it either by pulling the string vertically upward or along a frictionless inclined plane (see Figure 8). Which one of the following statements is true?
    (a) The magnitude of the tension force in the string is smaller in case (i) than in case (ii).
    (b) The magnitude of the tension force in the string is the same in both cases.
    (c) The work done on the block by the tension force is the same in both cases.
    (d) The work done on the block by the tension force is smaller in case (ii) than in case (i).
    (e) The work done on the block by gravity is smaller in case (ii) than in case (i).

Using the principle of conservation of mechanical energy, the correct response is (c). The most common incorrect responses were (d) and (e). The learning gain after instruction was very small on this item. Students had great difficulty focusing on the fact that since both blocks are raised by the same height at a constant speed, the work done by the gravitational force and tension force are the same in both cases according to the principle of conservation of energy. They got confused between the "force" and the

"work done by the force" and assumed that since it is easier to pull the block along the incline surface, there must be a smaller work done in case (ii). It is clear from student responses that they ignored the fact that the distance over which the force is applied is more along the incline surface than when it is pulled straight up.

Another case of difficulty in abstracting away from details comes from having to sum the abstract concept of energy across separate objects, in other words, reasoning about a system rather than individual parts. Consider problem 6, which was very difficult for students:

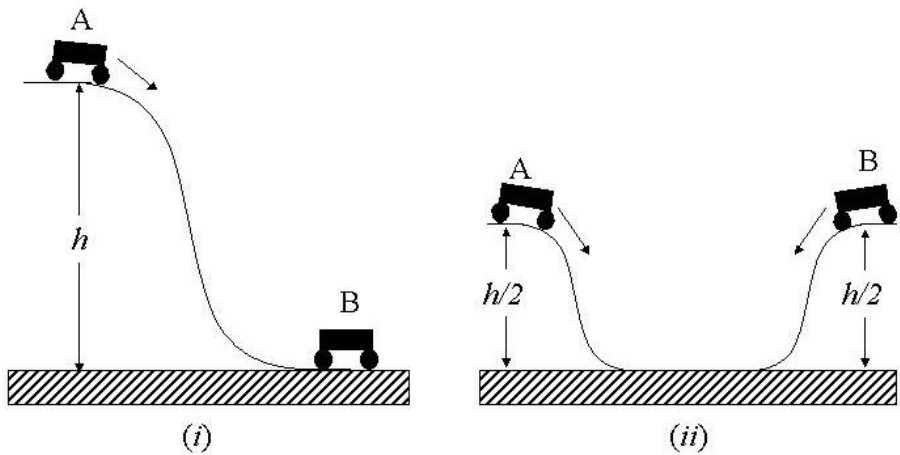

Figure 9. Carts A and B are identical in all respects before the collision. In scene (i): Cart A starts from rest on a hill at a height $h$ above the ground. It rolls down and collides "head-on" with cart B that is initially at rest on the ground. The two carts stick together. In scene (ii): Carts A and B are at rest on opposite hills at heights $h/2$ above the ground. They roll down, collide "head-on" with each other on the ground and stick together. Reprinted with permission from Singh (2003).

6. Which one of the following statements is true about the two-cart system shown in Figure 9 just before the carts collide in the two cases? Just before the collision on the ground,
   (a) the kinetic energy of the system is zero in case (ii).

   (b) the kinetic energy of the system is greater in case (i) than in case (ii).

   (c) the kinetic energy of the system is the same in both cases.

(d) the momentum of the system is greater in case (ii) than in case (i).

(e) the momentum of the system is the same in both cases.

Using the conservation of mechanical energy, the correct response for question (6) is (c). Unfortunately, the learning gain due to instruction was negligible. Students had similar difficulties both before and after instruction and all of the alternative choices were selected with almost equal frequency.

## *Confusion about Different Forms of Energy*

Our research shows that students often confuse different forms of energy, e.g., total mechanical energy, potential energy, kinetic energy, etc. This type of difficulty can make it difficult for students to be able to use the principle of conservation of energy appropriately. The response to question 7 illustrates this type of confusion (Lawson & McDermott, 1987; Singh, 2003; Van Heuvelen & Zou, 2001):

7.  Three bicycles approach a hill as described below:
(1) Cyclist 1 stops pedaling at the bottom of the hill, and her bicycle coasts up the hill.
(2) Cyclist 2 pedals so that her bicycle goes up the hill at a constant speed.
(3) Cyclist 3 pedals harder, so that her bicycle accelerates up the hill.

    Ignoring the retarding effects of friction, select all the cases in which the total mechanical energy of the cyclist and bicycle is conserved.
    (a) (1) only
    (b) (2) only
    (c) (1) and (2) only
    (d) (2) and (3) only
    (e) (1), (2) and (3)

The correct response to question 7 is (a) because in cases (2) and (3), the cyclist is using his internal energy to keep the bicycle moving at a constant speed or to accelerate it up the hill. Even after instruction, only 36% of the students provided the correct response.

The most popular distracters were (b) and (e). Individual interviews with students who selected option (b) shows that they felt that if the bicycle moves at a constant speed up the hill, the mechanical energy must be constant. What is unchanged in case (2) is the kinetic energy of the bicycle but the total mechanical energy is increasing since the potential energy increases. The students are confusing the kinetic energy for the total mechanical energy. Students who selected choice (e) thought that the only type of force that can violate the conservation of total mechanical energy is the frictional force. They ignored the internal energy of the person pedaling and assumed that in the absence of frictional forces, the total mechanical energy must be conserved. A student who chose (e) explained: *if you ignore the retarding effects of friction, mechanical energy will be conserved no matter what.* Other interviewed students who chose (e) also suggested that the retarding effect of friction was the only force that could change the mechanical energy of the system. While some students may have chosen (b) because they could not distinguish between the kinetic and mechanical energies, the following interview excerpt shows why that option was chosen by a student despite the knowledge that kinetic and mechanical energies are different:

> *S: I think it is (b) but I don't know... it can't be (c) because the person is accelerating.., that means (d) and (e) are not right...*
> *I: why do you think (b) is right?*
> *S: if she goes up at constant speed then kinetic energy does not change... that means potential energy does not change so the mechanical energy is conserved.., mechanical energy is kinetic plus potential.*
> *I: What is the potential energy?*
> *S: uhh... isn't it right?*
> *I: why is h not changing?*
> *S: (pause).. h is the height.. .1 guess h does change if she goes up the hill... hmm... maybe that means that potential energy changes. I am confused.. . .1 thought that if the kinetic energy does not change, then potential energy cannot change aren't the two supposed to compensate each other.... is it a realistic situation that she bikes up*

*the hill at constant speed or is it just an ideal case?*

The student is convinced that the mechanical energy is conserved when the bike goes up at a constant speed and he initially thinks that both the kinetic and potential energies must remain unchanged. When he confronts the fact that the potential energy is changing, instead of reasoning that the mechanical energy must be changing if the kinetic energy is constant, he thinks that it is probably not realistic to bike up the hill at a constant speed. He wonders if it is only possible in the idealized physics world. Although he ignores the work done by the non-conservative force applied on the pedal to keep the speed constant, his statements shed light on student's epistemological beliefs about how much one can trust physics to explain the everyday phenomena. A student who chose (c) (cases (1) and (2)) provided interesting explanation: *In case (1) the kinetic energy is transferred to potential energy so the mechanical energy is conserved and in case (2)... obviously. . if the speed is constant... mechanical energy is conserved. Case (3) is out because she is accelerating.* This example shows student's inconsistent thinking. There is acceleration not only in case (3) but also in case (1) (slowing down) but the student does not worry about it in case (1). At the same time, he put cases (2) and (3) in different categories although the cyclist was pedaling in both cases.

### *Difficulty with parametric dependence of energy on variables*

Students often have difficulty in determining the dependence of various forms of energy on different parameters that can make it difficult for them to apply energy principles appropriately. Student responses to question 8 illustrate this difficulty (Lawson & McDermott, 1987; Singh, 2003; Van Heuvelen & Zou, 2001):

8. You drop a ball from a high tower and it falls freely under the influence of gravity. Which one of the following statements is true?

(a) The kinetic energy of the ball increases by equal amounts in equal times.
(b) The kinetic energy of the ball increases by equal amounts over equal distances.
(c) There is zero work done on the ball by gravity as it falls.
(d) The work done on the ball by gravity is negative as it falls.
(e) The total mechanical energy of the ball decreases as it falls.

Students who chose (d) for question 8 believed that the work done by gravity on the ball falling from the tower is negative. Interviews show that many students did not invoke physics principles to come to this conclusion (e.g., the basic definition of work) but thought that the work must be negative if the ball is falling in the "negative y direction". Choices (a) and (b) were chosen with the same frequency. Students who chose the correct option (b) and the incorrect option (a) both knew that the kinetic energy of the ball increases as it falls. But the former group indicated that this increase was equal over equal distances (as the ball falls the potential energy decreases by equal amount over equal distances but the total mechanical energy is conserved) while the latter group indicated it was equal in equal times. Some students who chose the correct response used the process of elimination by noting that time has nothing to do with the conservation of mechanical energy. Students who focused on speed rather than kinetic energy were likely to get confused. The following is an excerpt from an interview with a student who chose (a) and started with a correct observation but then got mislead due to faulty proportional reasoning:

*I: Why do you think (a) is right?*
*S: Isn't it true that the velocity of the ball increases by like 9.8 m/s every second?... .kinetic energy is $(1/2)mv^2$ (writes down the formula) so it increases by equal amount over equal time.*
*I: Are you sure? Can you explain your reasoning?*
*S: I am pretty sure... (referring to the formula)... v increases by equal amount over equal times.. .so $v^2$ increases by equal amount over equal times... mass m is not changing...*

Student response to question 9 provides another example of reasoning about which parameters influence conservation of energy.

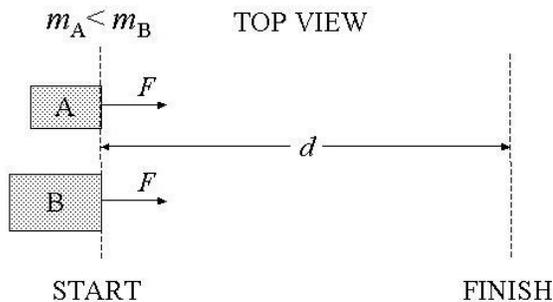

Figure 10. Two blocks are initially at rest on a frictionless horizontal surface. The mass $m_A$ of block A is less than the mass $m_B$ of block B. You apply the same constant force F and pull the blocks through the same distance d along a straight line as shown below (force F is applied for the entire distance d). Reprinted with permission from Singh.

9. Which one of the following statements about Figure 10 correctly compares the kinetic energies of the blocks after you pull them the same distance d?
    (a) The kinetic energies of both blocks are identical.
    (b) The kinetic energy is greater for the smaller mass block because it achieves a larger speed.
    (c) The kinetic energy is greater for the larger mass block because of its larger mass.
    (d) Not enough information, need to know the actual mass of both blocks to compare the kinetic energies.
    (e) Not enough information, need to know the actual magnitude of force F to compare the kinetic energies.

Question 9 has previously been investigated in-depth by McDermott et al. (Lawson & McDermott, 1987). Students have great difficulty in realizing that since identical constant forces are applied over the same distance to both masses (which start from rest), their kinetic energies are identical regardless of their masses. Only 29% indicated the correct choice (a) during the post-test and the strong distracters were (b) and (c). Interestingly, many students correctly stated that the velocity of block A will be greater but they had difficulty in reasoning beyond this. Interviews show that the choice

(b) was often dictated by the fact that the kinetic energy increases as the square of the speed but only linearly with mass (Lawson & McDermott, 1987).

McDermott et al. have developed and assessed tutorials (McDermott, Shaffer, & Physics Education Group, 2002) that significantly improve student understanding of energy concepts noted in the above examples.

We have developed some exploration problems that have been effective in improving student understanding of conservation of energy. One such exploration involves loop the loop demonstration with a ball and a track that looks like a roller coaster. One side of this track goes higher than the other side. Students are asked to predict various things such as the minimum height from which the ball should be released on higher side to be able to reach the end of the track on the lower side or the minimum height from which the ball should be released so as to complete a loop without losing contact with the track. In each case, students have to explain their reasoning and invoke the principle of mechanical energy conservation.

In sum, conservation of energy is a conceptually prominent in state science standards (although sometimes in quantitative form and sometimes in qualitative form), pivotal to the learning of physics, and yet very difficult for students to learn. We have identified some instructional problems that are useful for improving student learning in these areas.

## *Geometrical Optics*

Understanding of light and how it interacts with objects is important for all branches of science (Goldberg & McDermott, 1986, Goldberg & McDermott, 1987). Whether one is learning about microscopes, telescopes or human eye, one learns in geometrical optics that light travels in a straight line until it interacts with a material. After this interaction, the direction of light can change due to reflection, refraction, diffraction (which must be described by wave optics) and absorption. The Atlas of Science Literacy Project 2061 does not have a separate concept map for geometrical optics. However, according to the Atlas of Science Literacy Project 2061 "Waves" map (see Appendix C), children in grades 3-5 should be given an opportunity to learn about the basic properties of light. Helping students perform more in-depth qualitative analysis of wave phenomena in grades 6-8 can deepen this understanding. Quantitative analysis can be performed at the high school level in grades 9-12.

Geometrical and wave optics concepts (indicated in italics on the map in Appendix C) are basic to understanding a variety of phenomena pertaining to light. However, the basics of geometrical optics are surprisingly difficult even for students in college physics (Goldberg & McDermott, 1986, 1987).

The state science standards for CA, CO, RI, and WI have quite a varied treatment of optics. California and Rhode Island science standards provide the most direct reference to them, at both middle school and high school levels. Wisconsin standards (also at both middle school and high school levels) generally make reference to properties of light and models of them, which presumably must involve basic geometrical optics, although this must be inferred. Colorado science standards make the most indirect reference to this topic, with brief mention of light causing change in a system in the middle school standards, and some mention of analysis of characteristics of matter as they relate to emerging technologies such as photovoltaics.

Geometrical optics is in fact a cluster of related concepts that describe the rectilinear propagation of light in free space and its reflection and refraction when it interacts with matter (Goldberg & McDermott, 1986 and 1987; Wosilait, Heron, Shaffer, & McDermott, 1998). Teaching students about the properties of light is challenging and requires careful instructional planning. It has been documented that students have serious difficulties about the consequences of light traveling in a straight line in free space and getting reflected, refracted or absorbed after interacting with objects (Goldberg & McDermott, 1986 and 1987; Wosilait, Heron, Shaffer, & McDermott, 1998). The next three sections document the key difficulties that students have with propagation, reflection, and refraction of light.

## *Difficulty Understanding Propagation of Light Rays*

McDermott et al. (Wosilait, Heron, Shaffer, & McDermott, 1998) performed a study in which they investigated pre-service and in-service teachers and introductory physics students' understanding of light and shadow. They found that students had many common difficulties. They asked students to predict outcomes of experiments. After the prediction phase, students performed the experiments and tried to reconcile the differences between their prediction and observation.

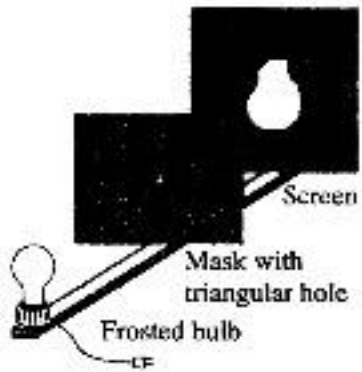

Figure 11. Experiment with frosted bulb shining through a pinhole and projecting onto a screen. Reprinted with permission from Wosilait, K., Heron, P., Shaffer , P. & McDermott, L. Development and assessment of a research-based tutorial on light and shadow. 1998, American Journal of Physics.

In one investigation (Wosilait et al., 1998), students were asked to predict what they would see on the screen when a mask with a very small triangular hole is placed between a broad extended source (a frosted bulb) and a screen (see Figure 11). This is a modified version of the classic pinhole camera setup in which a candle is placed in front of a mask with a very small circular hole and the image on the screen is an inverted candle. A very large number of students predicted that the screen will be lit only in the tiny triangular area in front of the hole in the mask. Even after performing the demonstration and observing an inverted image of the whole frosted bulb, many students could not reconcile the differences. This task turns out to be very difficult because of the way people generally interpret what it means for light to travel in a straight line (Wosilait, Heron, Shaffer, & McDermott, 1998). They do not realize that each point on the frosted bulb should be thought of as a point source of light that gives out light traveling in straight lines in all directions radially. In this study, many students seemed to believe that light can only travel horizontally through the triangular hole in the mask so that all of the light from the frosted bulb will be blocked except for the size of the hole.

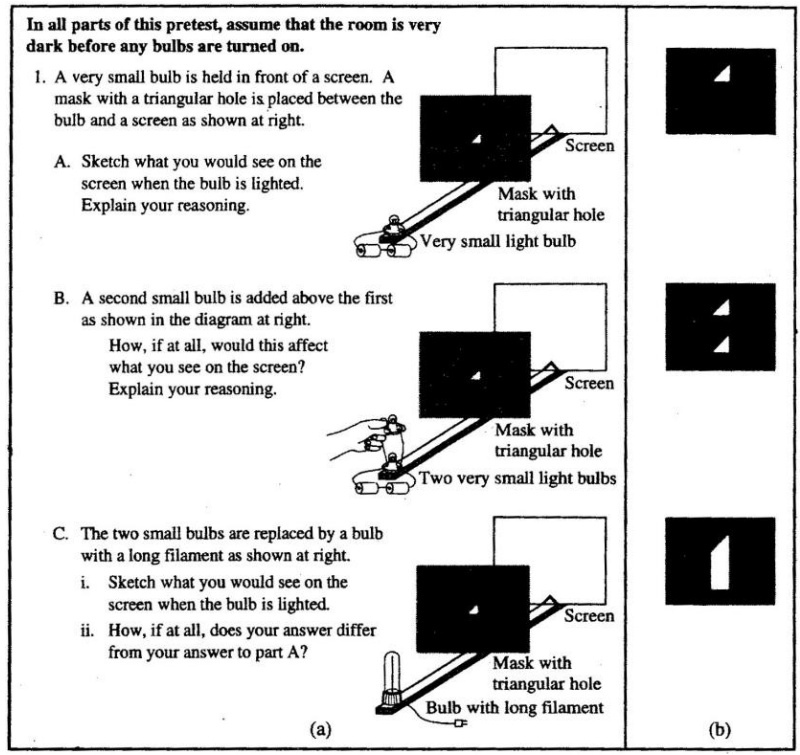

Figure 12. Reprinted with permission from Wosilait, K., Heron, P., Shaffer, P. & McDermott, L. Development and assessment of a research-based tutorial on light and shadow. 1998, American Journal of Physics.

In another investigation, McDermott et al. (1998) changed the relative sizes of the light source and the hole through which light passed before reaching the screen. This time the source of light was a very small bulb (about the size of a Christmas tree bulb) and the triangular hole was relatively large (see Figure 12). They asked students to predict what they would observe on the screen. If one correctly uses the fact that light travels in a straight line and the hole is much larger than the size of the source, one will come to the conclusion that the image on the screen will be triangular (the same shape as the hole). The size of the triangular image on the screen will change depending on the distance of the tiny bulb from the hole. Students were also asked to predict what will happen if there were two light bulbs, one underneath another in front of the same triangular hole. In this case, the bright image on the screen should be two triangles (the lower triangular image is formed by the upper bulb and vice versa). The last task in this set was a prediction of the image formed by a bulb with a long filament in front of the same triangular hole. These tasks turned out to be extremely difficult for most students because they had never carefully thought about what it means for light to travel in a straight line (Wosilait, Heron, Shaffer, & McDermott, 1998).

To help improve understanding of light and shadow, McDermott et al. (McDermott & Physics Education Group, 1996) developed and assessed laboratory-based, inquiry-oriented curriculum for pre-college teachers. They found that instructional materials that evolved from the iterative cycle proved effective in helping students understand the implications of the linear motion of light on the formation of shadow and images. In fact, after the modified curriculum, students were able to predict the type of image formed by complicated objects under diverse situations and the effect of the change of parameters such as the distance of the object or the screen from the hole.

Singh (2000, 2002) has developed several explorations that improve students' understanding of the concepts related to linear propagation of light and formation of images by reflection and refraction. For example, one exploration challenges students' pre-conceptions about shadows formed by obstacles including changes in the size of the shadow of the obstacle if the distance of the obstacle from the light source is increased. Another part of this exploration involves images formed by pinholes about which students have many common difficulties. These explorations have been found to be effective tools for helping students learn about rectilinear propagation of light and for developing confidence in drawing ray diagrams.

## *Difficulty with Reflection of Light*

Not only do students have difficulty in understanding the implications of the motion of light in a straight line from a source, they have difficulty understanding the formation of image by reflection of light from mirrors. Goldberg and McDermott (1986) performed a study in which they investigated student difficulties in understanding image formation by a plane mirror. The emphasis of their investigation was on examining the extent to which students connect formal concepts to real world phenomena. They found that most students can provide memorized answers to standard questions such as the image is the same distance behind the plane mirror as the object is in front. However, they cannot answer questions such as whether his/her distance from a small mirror would affect the amount he/she can see of his/her own image. Based upon interviews with many students, the researchers claim that even if traditionally taught students are given time and encouragement to reconsider, the students will most probably not even be able to draw a ray diagram that might help them answer such questions.

By gathering detailed information from interviews with a large number of students about four systematic tasks related to plane mirrors, Goldberg and McDermott (1986) were able to identify student difficulties in attempting to connect the principles of geometrical optics studied in class and the image they can see or imagine seeing in a real mirror. One difficulty that was common was the belief that an observer can see an image only if it lies along his or her line of sight to the object. Students who claimed that the object and the image were at equal distances from the mirror along the line of sight appeared not to be thinking that the mirror is a reflecting surface. In order to reconcile their experience of seeing an object shift with respect to the background, students sometimes introduced faulty parallax reasoning and predicted that an image would be in different positions for different observers.

Students also had great difficulty in deciding where, with respect to a ray diagram, the eye of an observer must be to see an image. Students often misinterpreted their past experiences. In trying to justify an incorrect prediction, students often provided reasoning that violated the law of reflection but students appeared to be unaware of it. Students have difficulty in understanding that a person can see something only if the light reflected from that object reaches the person's eyes (McDermott & Physics Education Group, 1996; McDermott, Shaffer, & Physics Education Group, 2002). This lack of understanding makes it very difficult to understand among other things, the phases of the moon. A common misconception is that the moon is always there in the sky but is sometimes covered by the clouds which gives rise to the different shapes.

McDermott et al. (McDermott & Physics Education Group, 1996) have developed an inquiry-based curriculum for K-12 teachers that is effective in helping dispel misconceptions about the phases of the moon. The curriculum helps K-12 teachers build a coherent understanding of the reflection of light and its implication for being able to see something.

### *Difficulty with Refraction of Light*

Formation of images by refraction of light is also quite challenging for students. Goldberg and McDermott (1987) performed a study in which they investigated student understanding of the real image formed by refraction through a converging lens or reflection through concave mirror. Students were often unable to apply the concepts and principles they had learned in their college introductory physics class to an actual physical system consisting of an object, a lens or a mirror, and a screen. Many students did not seem to understand the function of the lens, mirror or screen. The study included interviews in which students predicted outcomes of experiments, performed experiments and reconciled differences.

In one part of the study, students who had obtained a real image on the screen formed by the convex lens were asked to predict what would happen if the lens was removed. Since the lens forms the image, the image on the screen will disappear if the lens is taken away. Many students claimed that the image will become a little fuzzy if the lens is removed but remain on the screen nevertheless. Others claimed that the image will not be upside down anymore without the lens and will have the same orientation as the object. Even after performing the demonstration, many students did not know how to explain the disappearance of the image.

In another part of the study (Goldberg & McDermott, 1987), students were asked to predict what would happen on the screen to the real image that is formed by refraction through a convex lens if half of the lens was covered with a mask. Since each part of the lens forms the image, the image should remain on the screen but become half as intense. A large number of students claimed that image should get cut in half if half the lens was covered. Even after observing that the whole image remains intact but the image intensity decreases, most students could not draw ray diagrams to explain it.

Singh (2000, 2002) has developed an exploration with lenses that deals with the common incorrect assumption that covering half of a lens will cut the image in half or removing the lens will make the image fuzzy but the image will be present (in reality, if the image is formed by a lens, then removing the lens will make the image go away).

Students predict what will happen in these situations and then reconcile the difference between their prediction and observation. With the help of intensity measuring device (photocell), they find that covering half the lens reduces the intensity to half but since each part of the lens forms image, the full image remains. Using the ray diagram, students try to make sense of it. Students also notice that the image vanishes when the lens is removed.

Another exploration (Singh, 2000, 2002) deals with a model of human eye where students explore how the focal length of the eye changes in order to form a clear image on the retina. They learn about how defects in the eye prevent focal length of the eye from changing naturally to form a clear image on the retina. We have found that these explorations enhance student understanding of geometrical optics and help students build coherent knowledge structure where there is less room for misconceptions.

## *Summary*

In this paper, we have connected K-12 science standards in four states and maps of conceptual growth to research on student difficulties and research-based strategies for helping students related to three important physics concepts. These concepts are particularly important for educational interventions in the K-12 curriculum because they are pivotal on concept maps of related concepts that should be taught in the K-12 curriculum and students have many common difficulties in these areas. Since students who do not learn these concepts might have further difficulty in later learning; teachers must teach these concepts using research-based strategies keeping in mind the common difficulties students have in order to make significant progress.

These concepts are known to be extremely difficult for students and involve a variety of very robust misconceptions. Thus, learning these concepts is not a simple matter, is likely to require significant time in the curriculum, and must involve research-based curricula carefully constructed to help students build a robust knowledge structure. To aid teachers and curriculum developers in this work, we have provided an in-depth discussion of the core difficulties and some methods for challenging the alternative conceptions and helping students build a coherent knowledge structure.

One research-based strategy for helping students develop a solid grasp of these concepts discussed in this paper includes tutorials or guided inquiry-based learning modules, e.g., those developed by the University of Washington Physics Education Research group. Another strategy for helping students is to give them exploration problems that involve doing hands-on activities with guided worksheets that target common difficulties. These explorations start by asking students to predict what should happen in a particular situation, then asking them to perform the exploration and then reconcile the difference between their prediction and observation. It has been found that these research-based activities are more effective when students work on them in small groups.

Finally, we certainly do not wish to imply that other physics concepts are unimportant. However, we do wish to suggest that teachers, science education researchers, curriculum developers and those involved in the professional development of K-12 teachers and striving to improve K-12 education pay attention to the same three criteria (examining concept maps of different concept clusters, connecting these concepts to the state standards and exploring why these concepts are difficult and the research-based strategies that have been found effective in helping students in those areas) in deciding where to place emphasis on concepts to teach. For example, in some of the state standards, electricity and magnetism concepts were also prominent. It will be useful to perform a similar analysis for the electricity and magnetism concepts and connect the K-12 standards and maps of conceptual growth with the research on student difficulties and research-based strategies for helping students learn electricity and magnetism. The potential conceptual space is large, and progress should be made first on the concepts that are most important to overall student learning.

Table A1

California, Colorado, Rhode Island, and Wisconsin State Standards Related to Newton's Second Law

| Level | CA Science Content Standards for California Public Schools | CO Colorado Model Content Standards: Science | RI AAAS Project 2061: Benchmarks for Science Literacy | WI Wisconsin Model Academic Standards |
|---|---|---|---|---|
| Elementary | | | Grades 3-5 4.F. Motion Changes in speed or direction of motion are caused by forces. The greater the force is, the greater the change in motion will be. The more massive an object is, the less effect a given force will have. | |
| Middle | Grade 8 Focus on Physical Science 2e. Students know that when the forces on an object are unbalanced, the object will change its velocity (that is, it will speed up, slow down, or change direction). 2f. Students know the greater the mass of an object, the more force is needed to achieve the same rate of change in motion. | Grades 5-8 2.3 identifying and predicting what will change and what will remain unchanged when matter experiences an external force or energy change (for example, boiling a liquid; comparing the force, distance, and work involved in simple machines) | Grades 6-8 4.F. Motion An unbalanced force acting on an object changes its speed or direction of motion, or both. If the force acts toward a single center, the object's path may curve into an orbit around the center. | Grades 5-8 D.8.5 While conducting investigations, explain the motion of objects by describing the forces acting on them |
| *High* | Grades 9-12 Physics 1c. Students know how to apply the law F=ma to solve one-dimensional motion problems that involve constant forces (Newton's second law). 2f. Students know an unbalanced force on an object produces a change in its momentum. | Grades 9-12 2.3 describing and predicting …physical interactions of matter (for example, velocity, force, work, power), using word or symbolic equations | Grades 9-12 4. F. Motion The change in motion of an object is proportional to the applied force and inversely proportional to the mass. | Grades 9-12 D.12.7 Qualitatively and quantitatively analyze changes in the motion of objects and the forces that act on them and represent analytical data both algebraically and graphically |

Table A2

California, Colorado, Rhode Island, and Wisconsin State Standards Related to Conservation of Energy

|  | CA<br>Science Content Standards for California Public Schools | CO<br>Colorado Model Content Standards: Science | RI<br>AAAS Project 2061: Benchmarks for Science Literacy | WI<br>Wisconsin Model Academic Standards |
|---|---|---|---|---|
| Elementary |  |  |  |  |
| Middle |  |  | Grades 6-8<br>4.E. Energy Transformation<br>Energy cannot be created or destroyed, but only changed from one form into another. | Grades 5-8<br>D.8.7 While conducting investigations of common physical and chemical interactions occurring in the laboratory and the outside world, use commonly accepted definitions of energy and the idea of energy conservation |
| High | Grades 9-12 Physics<br>2. The laws of conservation of energy and momentum provide a way to predict and describe the movement of objects.<br>a. Students know how to calculate kinetic energy by using the formula $E=(1/2)mv^2$.<br>b. Students know how to calculate changes in gravitational potential energy near Earth by using the formula (change in potential energy) = mgh (h is the change in the elevation).<br>c. Students know how to solve problems involving conservation of energy in simple systems, such as falling objects.<br>e. Students know momentum is a separately conserved quantity different from energy. | Grades 9-12<br>2.2 identifying, measuring, calculating, and analyzing qualitative and quantitative relationships associated with energy transfer or energy transformation (for example, changes in temperature, velocity, potential energy, kinetic energy, conduction, convection, radiation, voltage, current).<br>2.3 observing, measuring, and calculating quantities to demonstrate conservation of matter and energy in chemical changes (for example, acid-base, precipitation, oxidation-reduction reactions), and physical interactions of matter (for example, force, work, power);<br>2.3 describing and explaining physical interactions of matter using conceptual models (for example, conservation laws of matter and energy, particle model for gaseous behavior). | Grades 9-12<br>4.E. Energy Transformation<br>Whenever the amount of energy in one place or form diminishes, the amount in other places or forms increases by the same amount. | Grades 9-12<br>D.12.10 Using the science themes, illustrate the law of conservation of energy during chemical and nuclear reactions |

Table A3

California, Colorado, Rhode Island, and Wisconsin State Standards Related to Optics

|  | CA<br>Science Content Standards for California Public Schools | CO<br>Colorado Model Content Standards: Science | RI<br>AAAS Project 2061: Benchmarks for Science Literacy | WI<br>Wisconsin Model Academic Standards |
|---|---|---|---|---|
| Elementary | Grade 3<br>2. b. Students know light is reflected from mirrors and other surfaces. |  |  |  |
| Middle | Grade 7: Focus on Life Science<br>6.f. Students know light can be reflected, refracted, transmitted, and absorbed by matter. | Grades 5-8<br>2.3 identifying and classifying factors causing change within a system (for example, force, light, heat) | Grades 6-8<br>4. F. Motion<br>Something can be "seen" when light waves emitted or reflected by it enter the eye—just as something can be "heard" when sound waves from it enter the ear.<br>Human eyes respond to only a narrow range of wavelengths of electromagnetic radiation-visible light. Differences of wavelength within that range are perceived as differences in color. | Grades 5-8<br>D.8.8 Describe and investigate the properties of light, heat, gravity, radio waves, magnetic fields, electrical fields, and sound waves as they interact with material objects in common situations |
| High | Grades 9-12 Physics<br>4f. Students know how to identify the characteristic properties of waves: interference (beats), diffraction, refraction, Doppler effect, and polarization. | Grades 9-12<br>2.3 relating their prior knowledge and understanding of properties of matter to observable characteristics of materials and emerging technologies (for example, semiconductors, superconductors, photovoltaics, ceramics) | Grades 9-12<br>4. F. Motion<br>Waves can superpose on one another, bend around corners, reflect off surfaces, be absorbed by materials they enter, and change direction when entering a new material. All these effects vary with wavelength. The energy of waves (like any form of energy) can be changed into other forms of energy. | Grades 9-12<br>D.12.9 Describe models of light, heat, and sound and through investigations describe similarities and differences in the way these energy forms behave. |

Appendix A: Conceptual Map of Force and Motion Concepts (adapted from Project 2061 Atlas of Science Literacy)

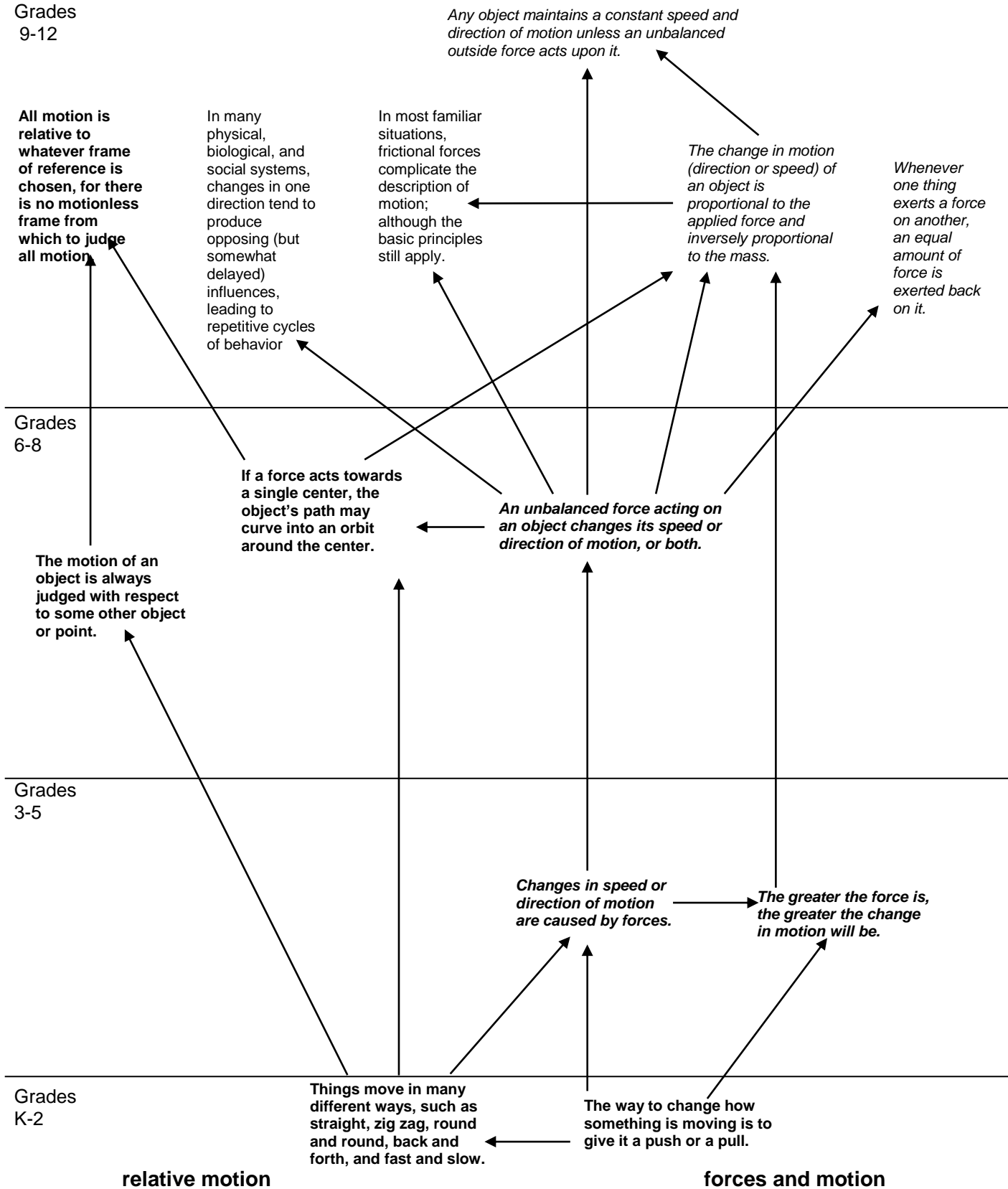

Appendix B: Conceptual Map of Energy Concepts

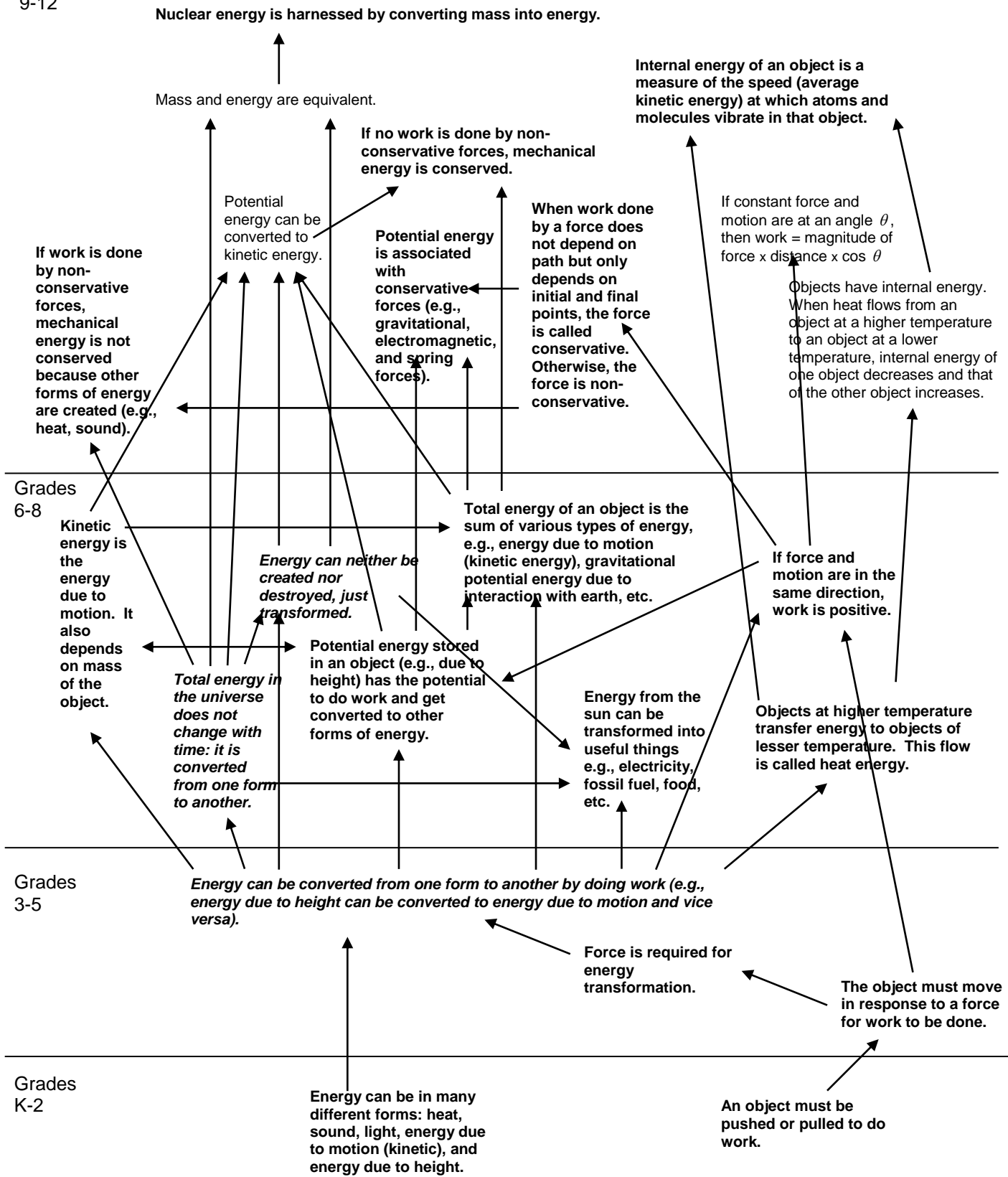

Appendix C: Conceptual Map of Waves Concepts (adapted from Project 2061 Atlas of Science Literacy)

**Grades 9-12**

**All motion is relative to whatever frame of reference is chosen, for there is no motionless frame from which to judge all motion.**

The observed wavelength of a wave depends upon the relative motion of the source and the observer. If either is moving toward the other, the observed wavelength is shorter; if either is moving away, the wavelength is longer.

Accelerating electric charges produce electromagnetic waves around them. A great variety of radiations are electromagnetic waves: radio waves, microwaves, radiant heat, visible light, ultraviolet radiation, x rays, and gamma rays. These wavelengths vary from radio waves, the longest, to gamma rays, the shortest. In empty space, all electromagnetic waves move at the same speed—the "speed of light."

*Waves can superimpose on one another, bend around corners, reflect off surfaces, be absorbed by materials they enter, and change direction when entering a new material. All of these effects vary with wavelength.*

In many physical, biological, and social systems, changes in one direction tend to produce opposing (but somewhat delayed) influences, leading to repetitive cycles of behavior.

The energy of waves (like any form of energy) can be changed into other forms of energy.

**Grades 6-8**

**Human eyes respond only to a narrow range of wavelengths of electromagnetic waves—visible light. Differences of wavelength within that range are perceived as differences of color.**

**Wave behavior can be described in terms of how fast the disturbance spreads, and in terms of the distance between successive peaks of the disturbance (the wavelength).**

**Light from the sun is made up of a mixture of many different colors of light, even though to the eye the light looks almost white. Other things that give off or reflect light have a different mix of colors.**

**Something can be "seen" when light waves emitted or reflected by it enter the eye—just as something can be "heard" when sound waves**

**Light acts like a wave in many ways. And waves can explain how light behaves.**

**Vibrations in material set up wavelike disturbances that spread away from the source. Sound and earthquake waves are examples. These and other waves move at different**

**Grades 3-5**

*Light travels and tends to maintain its direction of motion until it interacts with an object or material. Light can be absorbed, redirected, bounced back, or allowed to pass through.*

**One way to make sense of something is to think how it is like something more familiar.**

**How fast things move differs greatly.**

**Grades K-2**

**Things that make sound vibrate.**

**Things move in many different ways, such as straight, zig zag, round and round, back and forth, and fast and slow.**

**light**    **wave motion**    **vibrations**